\documentclass{svjour3}                     

\usepackage[numbers]{natbib}
\RequirePackage{fix-cm}

\usepackage{Shortcuts}
\usepackage{Symbols}
\usepackage{Text}

\usepackage{graphicx}

\usepackage{epsfig}
\usepackage{subfigure}

\def \MZIsys {MZI$_\text{sys}$}
\def \MZIdet {MZI$_\text{det}$}

\journalname{Quantum Stud.: Math. Found.}

\begin{document}

\title{How to extract weak values from a mesoscopic electronic system}

\author{Iliya Esin \and Alessandro Romito \and Yuval Gefen}

\institute{I.Esin \at Department of Condensed Matter Physics, The Weizmann Institute of Science, Rehovot 76100, Israel
           \and
           A. Romito \at Department of Physics, Lancaster University, Lancaster, LA1 4YB, United Kingdom
		   \and
		   Y. Gefen \at Department of Condensed Matter Physics, The Weizmann Institute of Science, Rehovot 76100, Israel}

\date{Received: date / Accepted: date}

\maketitle

\begin{abstract}
Weak value  (WV) protocols may lead to extreme  expectation values that are larger than the corresponding orthodox expectation values. Recent  works have proposed to implement this concept in nano-scale electronic systems. Here we address the issue of how one calibrates the setup in question, maximizes the measurement's sensitivity, and extracts distinctly large WVs. Our concrete setup consists of two Mach-Zehnder interferometers (MZIs): the ``system'' and the ``detector''. Such setups have already been implemented in experiment.
\end{abstract}

\section{Introduction and motivation}
\label{intro}
Strong measurement in quantum mechanics leads to the collapse of the measured system's wave function \cite{Neumann1955}. The challenge of performing a non-invasive measurement is interesting both from the view point of foundations of quantum mechanics, and  for concrete applications (e.g. quantum computation). Weakly measuring an observable, while weakly disturbing the system, provides only partial information on the state of the latter. Weak measurements, due to their backaction, can be exploited for quantum feedback schemes \cite{Zhang2005,Vijay2012,Wiseman2010} and conditional measurements. The latter entail WV protocols \cite{Aharonov1988} and derivatives thereof.

A standard two-step WV protocol consists of a weak measurement step  (of the observable $A$), followed by a strong one (of $B$), $\com{A}{B}\ne0$. The outcome of the first is conditional on the result of the second (postselection), \ti{i.e.}, one admits the outcome of the weak measurement of $A$ provided the result of the (strong) measurement of $B$ coincides with a prescribed value, $\av{A}_{WV}=\frac{\tr{A\cdot \Pi_B}}{\tr{\Pi_B}}$, with $\Pi_B$ being the projection operator on the postselected subspace. WVs have been observed in experiments \cite{Ritchie1991,Pryde2005,Hosten2008,Groen2013}. Their unusual expectation values \cite{Aharonov1988} may be utilized for various purposes, including weak signal amplification \cite{Dixon2009,Starling2009,Brunner2010,Starling2010,Dressel2014,Zilberberg2011}, quantum state discrimination \cite{Lundeen2011,Zilberberg2013}, and non-collapsing observation of virtual states \cite{Romito2014}. There have been recent proposals to implement WV protocols in the context of solid state systems \cite{Shpitalnik2008,Romito2008,Dressel2012}. Apart from the realization with superconducting qubits in resonant cavity\cite{Groen2013}, setups made up of two (electrostatic interaction coupled) MZIs, operating in the quantum Hall regime, are of particular importance, given their immediate experimental availability \cite{Ji2003,Weisz2014} and their versatile and controllable nature. In such setups one MZI plays the role of ``the system'', the other being ``the detector''.

Shpitalnik \etal \cite{Shpitalnik2008} have  implemented, in  principle, a WV  protocol in this setup, and have shown that the outcome of such measurement may produce a complete tomography of the WV  measured in the system's MZI. An exhaustive analysis of the correlated signal in this system has been reported in the single particle regime\cite{Dressel2012}, and the many body effects on the weak to strong measurement crossover have been classified \cite{Esin2014}. The fact that such a  protocol is amenable to experimental verification \cite{Weisz2014} has been undermined by the lack of a concrete manual on how to implement it.

The present analysis is meant to point out that WVs in solid state physics setups are feasible. We point out, \ti{vis-\`a-vis} a two MZI setup, which measurements need to be performed, how the sensitivity of the protocol can be controlled, how the readings of the detector should be calibrated, and consequently-- how extreme WVs can be obtained and identified.

\begin{figure}
	\includegraphics[width=8.6cm]{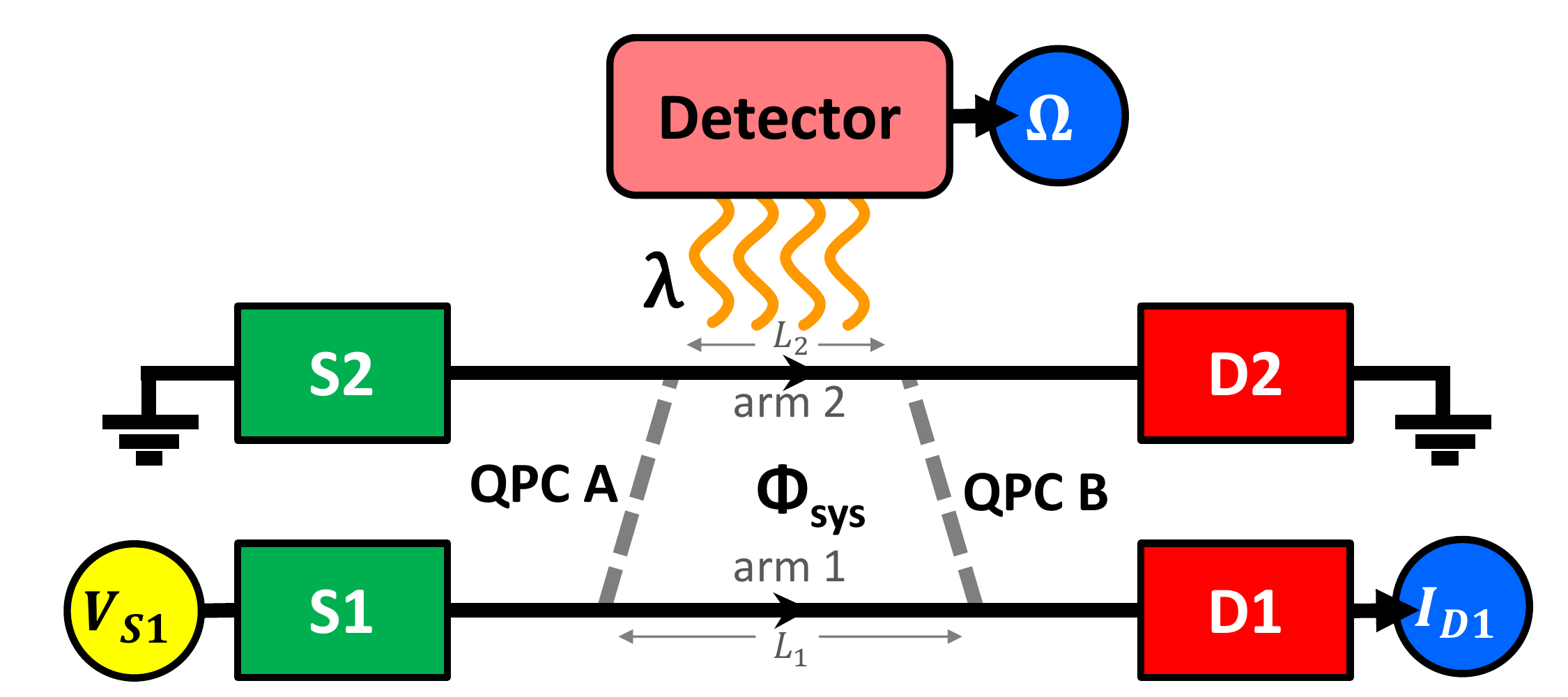}\\
	\caption{(Color online) A weakly detected MZI. The average reading of the detector, $\W$, is linearly proportional (to leading order in system--detector coupling, $\lm$) to the total electric charge on arm 2 of the interferometer at the time of the measurement. (If the detector is a current--carrying quantum point contact (QPC), then $\W(t)$ is equal to $\W(t)\eqs\int_{t-\ta_{fl}/2}^{t+\ta_{fl}/2} I(t') dt'$, where $I(t)$ is the current through the QPC and $\ta_{fl}=L_2/v$, with $L_2$ being the length of arm 2 and $v$ the velocity of the chiral edge mode). The strong follow-up measurement detects the charge (current pulse) arriving at the drain $D1$. Registering the outcome of the first measurement is subsequently conditioned on detecting a `click' in $D1$. This is the WV of the charge  in arm 2. \label{fig:SetupPrinciple}}
\end{figure}

\section{The system: a MZI}
Our system consists of a MZI depicted schematically in Fig.~\ref{fig:SetupPrinciple}. Electrons are injected from the source contact $S1$, kept at a finite voltage bias $V_{S1}$, and are collected at drain $D1$. The other terminals of the interferometers are grounded. The Quantum Point Contacts (QPCs) A and  B allow tunnelling of the electrons between arms $1$ and $2$ of the system; the electrons collected at $D1$ are the result of the interference  of different electron trajectories, and are sensitive to the magnetic flux $\Phi_{\rm sys}$.
A detector is electrostatically coupled to the charge in arm $2$ of the interferometer.
At this stage we refer to a general detector, weakly coupled to the system.

We consider explicitly the case where the bias current fed into the MZI is diluted (for example, one modifies the setup depicted in Fig.~\ref{fig:SetupPrinciple} such that most electrons emitted from the source $S1$ are backscattered before arriving in the MZI). In that case, the width of an electron's wave packet is much smaller than the distance between two consecutive electrons.
Moreover, we require that the time separation between successive injections of non-equilibrium electrons ($\ta_{V_{S1}}\eqd 2\p \hb/eV_{S1}$
is much larger than the electron's time-of-flight through the interferometer's arm\cite{Youn2008}: $\ta_{V_{S1}}\gg\ta_{fl}$. To reduce adverse decoherence effects one may consider the limit of low temperature, low voltage bias, and nearly symmetric interferometers (i.e., nearly equal arm lengths). The conditions are met in actual experiments \cite{Ji2003,Weisz2014}.

The first step of our protocol consists of weakly measuring the electric charge $Q_2$, flowing through arm 2 of the interferometer. This (weak) measurement is performed as a snapshot at time $t_W$ of the electric charge along arm 2.  The follow-up (strong) measurement detects the charge arriving at the drain $D1$ with a delay, $t_{delay}$, due to the finite propagation time of the charge from the weak detector to $D1$. The measurement itself consists of integrating the current pulse  over a window of time, $\bS{t_W+t_{delay}, t_W+t_{delay}+\ta_{fl}}$, which corresponds  to the time of flight of an electron through arm 2 (the latter is of length $L_2$; $\ta_{fl} = L_2/v$, where $v$ is the Fermi velocity of the non-equilibrium electrons). We denote this integrated  current by $\ta_{fl}I_{D1}$. Under the conditions of diluted injected current specified above, the postselection signal, $I_{D1}$, can reveal the detection of one or no electrons collected at $D1$, we then condition the acceptance of the first measurement of $Q_2$ on a `click' in $D1$. Therefore the weak value of the charge on arm 2 is
\begin{equation}
\av{Q_2}_{WV}=\frac{\av{Q_2(t_W)\cdot I_{D1}(t_W+t_{delay}+\half\ta_{fl})}}{\av{I_{D1}}}.
\label{eq:WVdefinition}
\end{equation}
We now relate the abstract WV defined above to a measurable quantity. We assume that (to leading order in system--detector coupling) the average signal of the detector is linearly proportional to measured charge, \ti{i.e.},  $\av{\dl \W}=\cS \av{Q_2}$, where $\W$ is the signal of the detector with $\dl \W\eqd \W-\W\at{\av{Q_2}=0}$ and $\cS$ is the sensitivity of the detector defined as
\begin{equation}
\cS\eqd \frac{\dpa \av{\W}}{\dpa \av{Q_2}}.
\label{eq:DefinitionSensitivity}
\end{equation}
We define the measured $\av{Q_2}^M_{WV}$ as
\begin{equation}
\av{Q_2}^M_{WV}\eqd\frac{1}{\cS}\frac{\av{\dl \W\cdot I_{D1}}}{\av{I_{D1}}}.
\label{eq:WVdefinitionDet}
\end{equation}
We expect the latter to be proportional to the WV $\av{Q_2}_{WV}$ (c.f Eq.~\eqref{eq:WVdefinition}).

\begin{figure*}
	\subfigure[two][]{\includegraphics[width=8.6cm]{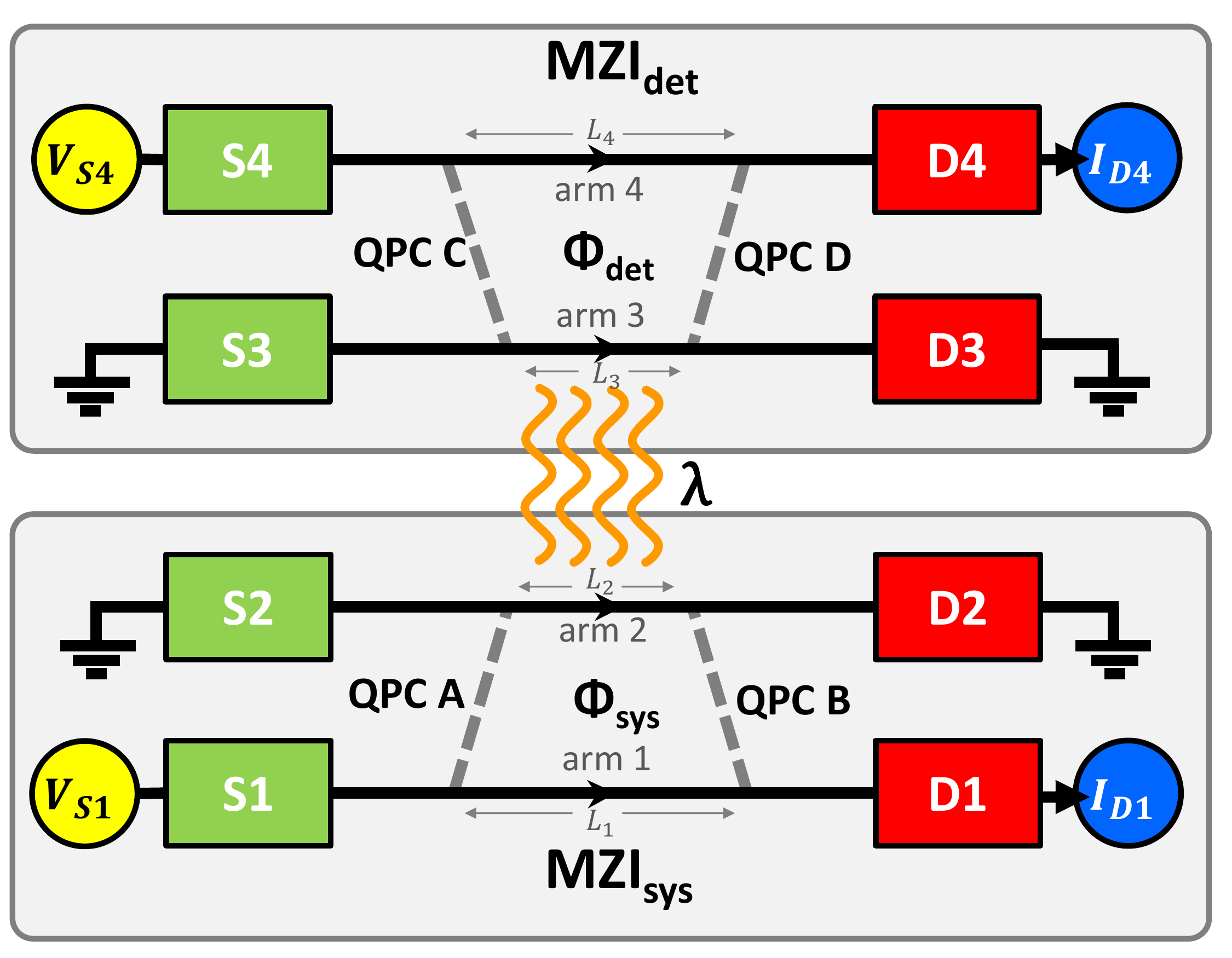}
		\label{subfig:SetupSchematic}}
	\subfigure[one][] {\includegraphics[width=8.6cm]{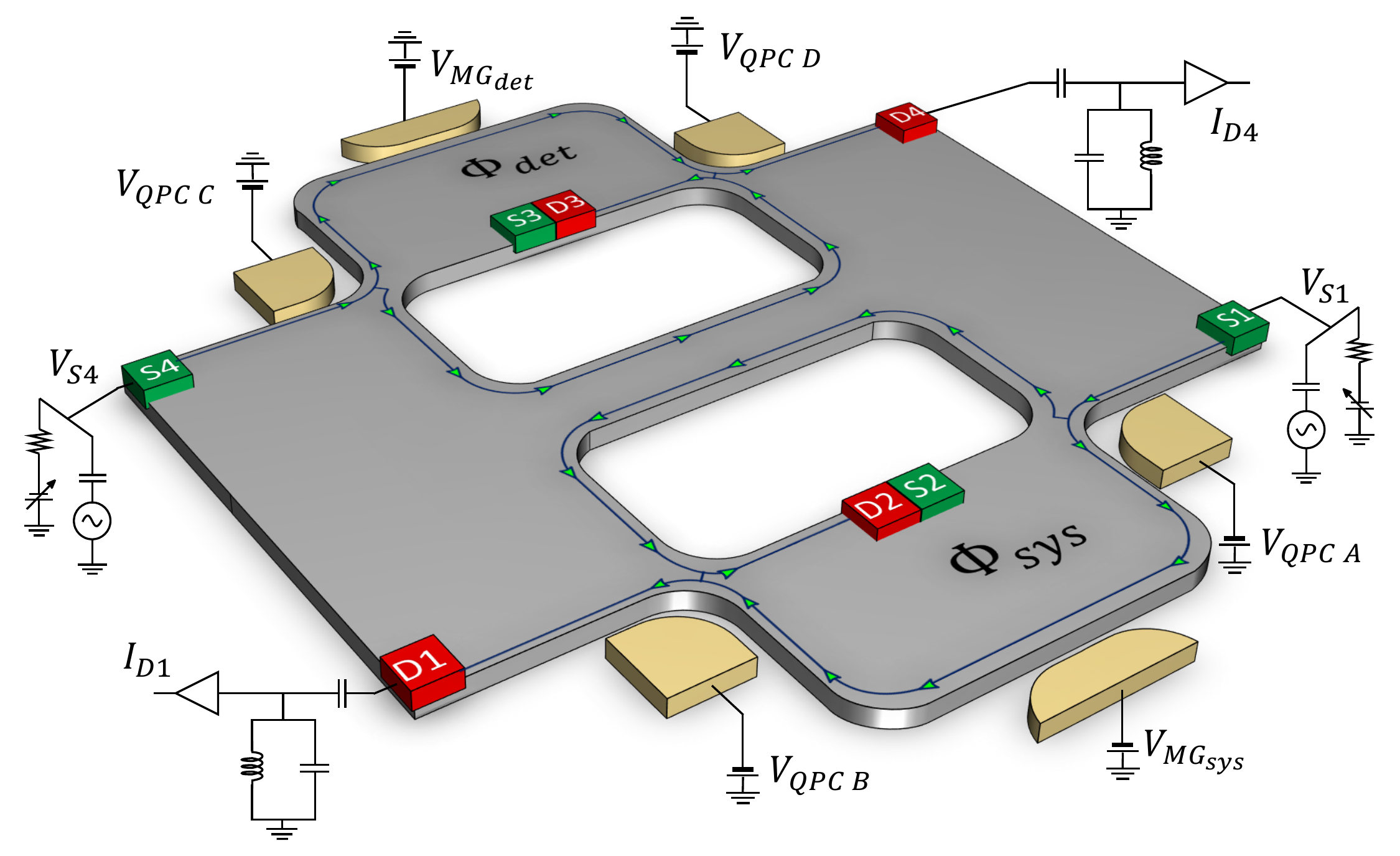}
		\label{subfig:SetupReal}}
	\caption{(Color online) A concrete realization of the setup. (a) A schematic layout. (b) A realistic design. 
		The system and the detector are realized by \MZIsys and \MZIdet, which are weakly (with strength $\lm$) electrostatically coupled through arms 2 and 3.
		The sources $S1$ and $S4$ are biased with the voltages $V_{S1}$ and $V_{S4}$ respectively, while the signals are measured in $D1$ and $D4$. 
		The contacts $S2$,$D2$,$S3$ and $D3$ are connected to the ground. The gate voltages $V_{\text{QPC A-D}}$ control the inter-arm tunneling amplitudes of electrons near the respective QPC. 
		The modulation gate biases $V_{MG_{sys}}$ ($V_{MG_{det}}$) control the effective magnetic fluxes $\F_{sys}$ ($\F_{det}$) by modifying the areas encircled by electronic trajectories along the device.\label{fig:Setup}}
\end{figure*}

\section{The detector: a MZI}
We realize the detector by \MZIdet as is shown in Fig. ~\ref{fig:Setup}. Arm 3 of the detector is electrostatically weakly coupled to system's arm 2.  We assume that the detector MZI is tuned, analogously with the system MZI, to have a diluted incident current and a time-separation between incident electrons larger than the time-of-flight across the interferometer. The current of the detector integrated over $\ta_{fl}$, $\ta_{fl}I_{D4}$, is sensitive to a charge on arm 2, and serves as a \ti{pointer variable} --- It plays the role of $\W$ in the general formulation of the previos section.  The (weak) signature of the system--detector interaction is a small  additional  phase gain of the wavefunction when a pair of electrons flow simultaneously along the arms 2 and 3 respectively\cite{Dressel2012,Neder2007}.

Therefore, the definition of $\av{Q_2}^M_{WV}$ (Eq.~\eqref{eq:WVdefinitionDet}) in the case of MZI as a detector (cf. Fig.~\ref{fig:Setup}) reads
\begin{equation}
\av{Q_2}^M_{WV}=\frac{1}{\cS}\bS{\frac{\av{I_{D4}I_{D1}}}{\av{I_{D1}}}-\av{I_{D4}}\at{00}}.
\label{eq:WVdefinitionMZI}
\end{equation}
It requires the measurement of the current--current correlator $\av{I_{D4}I_{D1}}$, the average current in $D1$, $\av{I_{D1}}$, and the average current in $D4$ when the transmission of QPC A is set to zero, $\av{I_{D4}}\at{00}$. The value of $\cS$ is not known. It is evidently an essential element to determine whether the outcome of a weak value protocol in an experiment yields an anomalous (large) value. In order to determine $\cS$, or equivalently to know the boundaries of the detector signal for unconditioned measurements, one needs a suitable calibration and characterization of the detector's sensitivity.

\begin{figure}
	\includegraphics[width=8.6cm]{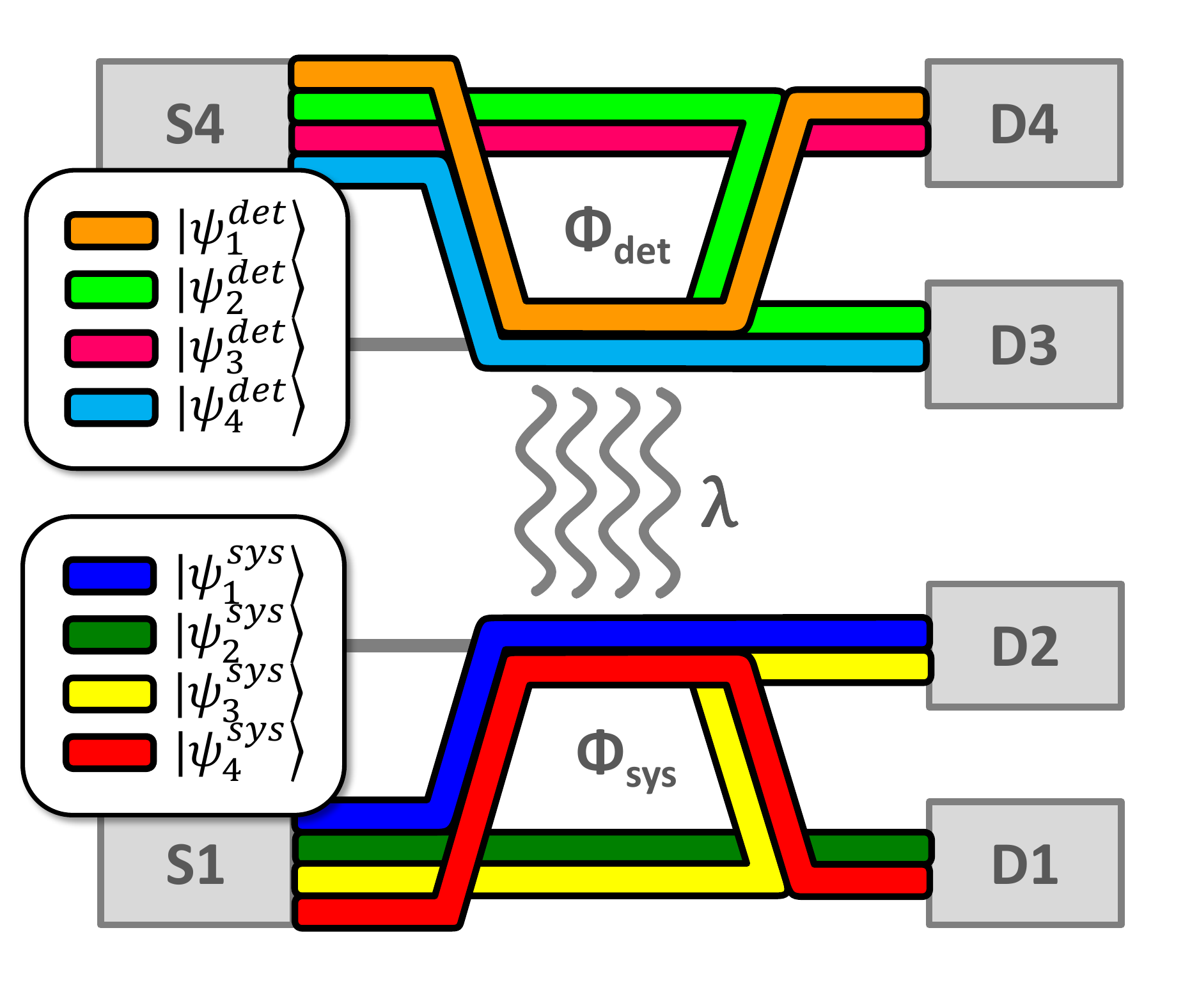}\\
	\caption{(Color online) Electronic trajectories along MZIs. The underlying scattering electronic wavefunction can be constructed as an entangled product of electronic trajectories in \MZIsys and in \MZIdet. Those are depicted schematically by different colors. cf. Appendix~\ref{sec:ExplicitTrajectories} for explicit expressions for the electronic trajectories.\label{fig:Paths}}
\end{figure}

\section{The system-detector coupling}
\label{modello}

To gain physical insight on the detector's response and determine its proper calibration we resort to a two-particle picture (one electron passing through \MZIsys and another in \MZIdet). The model is a valid description of the interaction between the system and detector electrons in the regime of diluted electron currents we are considering.


The two-particle scattering state, $\ket{\Y_{sys,det}}$, near the drains (after the QPCs B and D) can be expressed in terms of partial electronic trajectories (cf. Fig.~\ref{fig:Paths}) as,
\begin{equation}
\ket{\Y_{sys,det}}=\sum_{ij}e^{i\lm_{ij}}c^{sys}_i\ket{\y^{sys}_i} c^{det}_j\ket{\y^{det}_j},
\label{eq:PartialPaths}
\end{equation}
where $c^{sys}_{i}$ ($c^{det}_{i}$) are amplitudes of the trajectories through \MZIsys (\MZIdet) omitting the coupling between the interferometers (cf. Appendix \ref{sec:ExplicitTrajectories}), and $\lm_{ij}=\condf{\lm, &(i=1\lor 4)\land (j=1\lor 4)\\0,&o.w.}$ is the weak coupling term ($\lm\ll1$).
The WV of $Q_2$  (cf. Eq.~\eqref{eq:WVdefinition}) may be expressed in terms of the amplitudes $c_i$ as
\begin{equation}
\av{Q_2}_{WV}=\av{Q_2}_0\frac{c^{sys}_4}{c^{sys}_4+c^{sys}_2}.
\label{eq:Q2WVDefinition}
\end{equation}
where $\av{Q_2}_0\eqa\frac{e^2V_{S1}L_2}{2\p \hb v}$ is the average excess charge on the segment $L_2$. Similarly, we define
\begin{equation}
\av{Q_3}_{WV}=\av{Q_3}_0\frac{c^{det}_1}{c^{det}_1+c^{det}_3},
\label{eq:Q3WVDefinition}
\end{equation}
with $\av{Q_3}_0\eqa\frac{e^2V_{S4} L_3}{2\p \hb v}$, as the weak value of the charge on arm 3 conditioned on a signal in $D4$. With these definitions the signal in $D4$ to first order in $\lm$ is given by (cf. Appendix~\ref{sec:DerivationCorrelators})
\begin{equation}
\av{I_{D4}}=\av{I_{D4}}_0\bR{1+2\tilde\lm\im{\av{Q_3}_{WV}\bS{\av{Q_2}+\av{Q_2^{bg}}}}},
\label{eq:DetectorCurrent}
\end{equation}
where
$\tilde\lm=\lm/\av{Q_2}_0\av{Q_3}_0$ and
$\av{Q_2}=\av{Q_2}_0\bR{\abs{c^{sys}_1}^2+\abs{c^{sys}_4}^2}$
is the out-of-equilibrium and $\av{Q_2^{bg}}$ is the background charge on arm 2, $\av{I_{D4}}_0=\bR{\av{Q_3}_0/\ta_{fl}}\abs{c^{det}_1+c^{det}_3}^2$ is the current measured at $D4$ in the absence of interaction ($\lm=0$).
We obtain also the explicit expression for Eq.~\eqref{eq:WVdefinitionMZI} (cf. Apeendix~\ref{sec:DerivationCorrelators}), given by
\begin{equation}
\av{Q_2}^M_{WV}=\frac{2\tilde\lm\av{I_{D4}}_0}{\cS}\im{\av{Q_2}_{WV}\bR{\av{Q_3}_{WV}-\av{Q_3}}}.
\label{eq:WVexplicit}
\end{equation}

From the analysis it also follows that the sensitivity of the MZI detector (cf. Eq.~\eqref{eq:DefinitionSensitivity}) is
\begin{equation}
\cS=2\tilde\lm\av{I_{D4}}_0\im{\av{Q_3}_{WV}},
\label{eq:ExplicitSensitivity}
\end{equation}
with the explicit expression $\cS=2\av{Q_3}_0^2\tilde\lm/\ta_{fl}\abs{c^{det}_1c^{det}_3}\sin{\tilde\f}$. Here $\tilde\f$ is the total phase difference between the trajectories $\ket{\y^{det}_1}$ and $\ket{\y^{det}_3}$ (cf. Fig.~\ref{fig:Paths}), including  contributions from the AB flux, the orbital phase, and impurity scattering.
As follows from eq.~\eqref{eq:ExplicitSensitivity}, the maximal sensitivity is obtained when (i) the QPCs C and D are set to half transmission and (ii) the total phase difference  is $\tilde\f=\p/2+\p n$. The first requirement may be achieved by individually adjusting the gate voltages of the QPCs. To set the phase $\tilde\f$ of the detector to the maximal sensitivity point we employ a calibration protocol discussed below.

\begin{figure}
	\includegraphics[width=8.6cm]{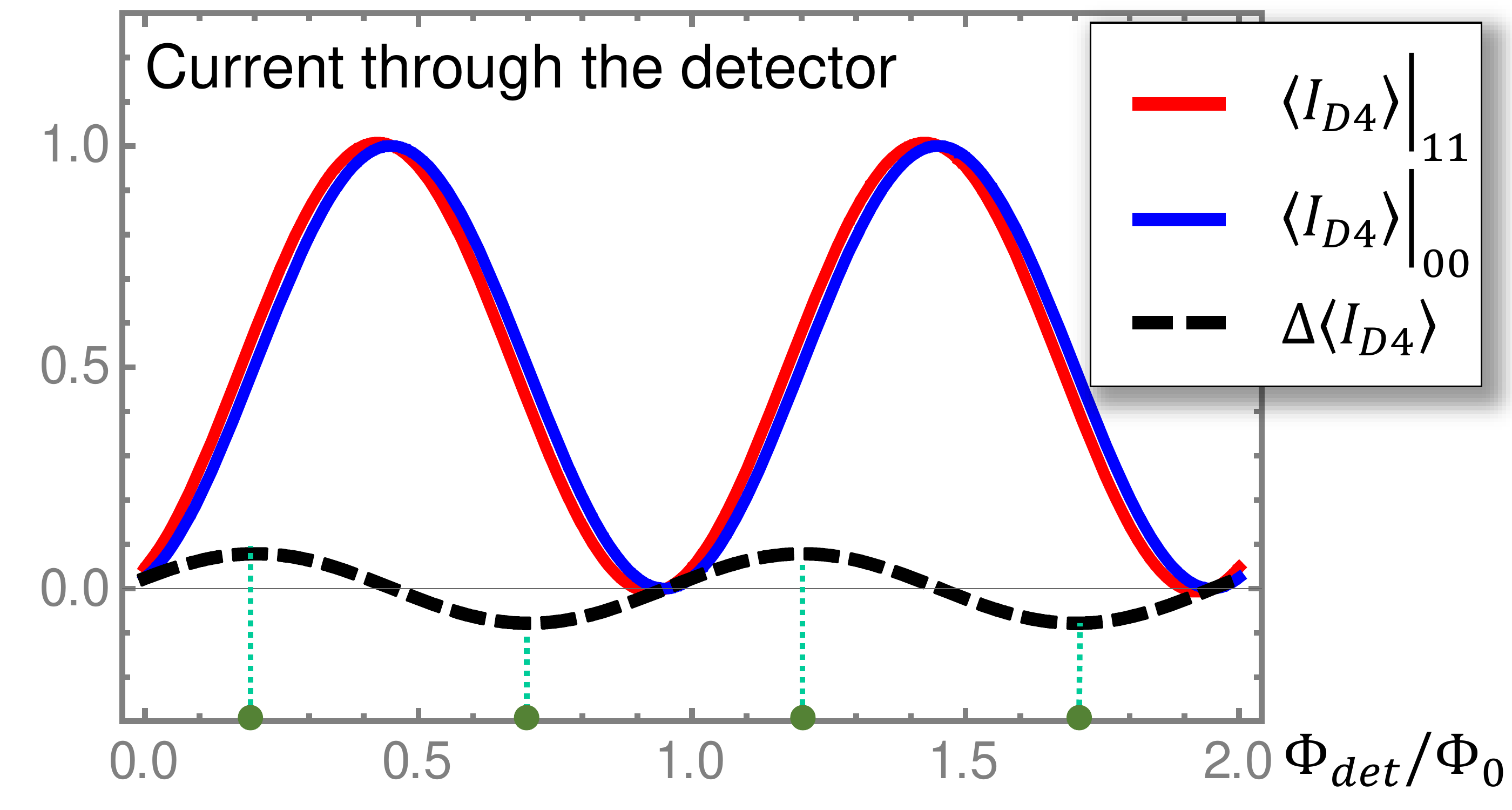}\\
	\caption{(Color online) Calibration of the detector. The QPCs C and D are set to half transmission ($\abs{t_C}=\abs{t_D}=1/\sqrt{2}$), $\tilde\lm=0.05\p$. The currents are measured in units of $\av{Q_3}_0$. The red (blue) curve corresponds to the signal in $D4$ when both QPC A and QPC B are fully transmitting (reflecting), $\av{I_{D4}}\at{11}$ ($\av{I_{D4}}\at{00}$). The black dashed curve is their difference: $\D\av{I_{D4}}\eqd \av{I_{D4}}\at{11}-\av{I_{D4}}\at{00}$. The detector is maximally sensitive at the extremas of $\D\av{I_{D4}}$. Those points are highlighted by green dots. \label{fig:CalibrationPlot}}
\end{figure}

\section{Calibration of the detector and extraction of weak values}
\label{sec:Calibration}

Let us  assume the QPCs are already tuned to the half transmission point. Here we present a calibration protocol for the phase $\tilde\f$ governing the interference signal registered in the detector.
In the first step of this protocol, one sets QPC A of the system(!) (by tuning the gate voltage $V_{\text{QPC A}}$) to `full reflection' (no current through both QPC A and QPC B \footnote{In fact, one can perform this calibration employing a general value of QPC B transmissivity.}, cf. Fig.~\ref{fig:Setup}), and measures the current $\av{I_{D4}}\at{00}$ as function of the detector's flux $\F_{det}/\F_0$, where $\F_0=h/e$ is the magnetic flux quantum. Evidently the signal is independent of the system's flux $\F_{sys}$, and of the transmission of QPC B. In this configuration the amplitudes for trajectories $\ket{\y_1^{sys}}$ and $\ket{\y_4^{sys}}$ vanish and consequently $\av{Q_2}=0$. In the second step both QPC A and QPC B are tuned to the opposite limit of `full transmission' (a charge arriving from $S1$ is deflected, with probability 1, to arm 2 (cf. Fig.~\ref{fig:Setup})), setting the weight of the trajectories $\ket{\y_2^{sys}}$ and $\ket{\y_3^{sys}}$ to zero. In this configuration the current in \MZIsys flows through arm 2 and the charge $\av{Q_2}$ reaches its maximal value, $\av{Q_2}_{max}$. In this limit, the current measured in the detector is denoted  $\av{I_{D4}}\at {11}$.

A representative plot of $\av{I_{D4}}$ for the two tunings: $\av{I_{D4}}\at{00}$ and $\av{I_{D4}}\at{11}$ respectively, and their difference $\D\av{I_{D4}}\eqd\av{I_{D4}}\at{11}-\av{I_{D4}}\at{00}$ is shown in Fig.~\ref{fig:CalibrationPlot} as a function of magnetic flux $\F_{det}$. The maximal sensitivity of the detector is achieved when one sets the magnetic flux to extremal points of $\D\av{I_{D4}}$. Values of the latter are depicted in the figure.

The calibration process prescribes the  values of the sensitivity $\cS$ (cf. Eq.~\eqref{eq:ExplicitSensitivity}), $\av{Q_3}_{WV}$ (cf. Eq.~\eqref{eq:Q3WVDefinition}, $\av{I_{D4}}_0$ and $\av{Q_3}=\av{Q_3}_0\bR{\abs{c_1^{det}}^2+\abs{c_4^{det}}^2}$ (cf. equations following Eq.\eqref{eq:DetectorCurrent}). It turns out that the calibration at maximal sensitivity yields $\re{\av{Q_3}_{WV}}=\av{Q_3}$.
Using the latter equality and Eq.~\eqref{eq:ExplicitSensitivity} we rewrite Eq.~\eqref{eq:WVexplicit} as
\begin{equation}
\av{Q_2}^M_{WV}=\re{\av{Q_2}_{WV}}\text{ (at maximal sensitivity)},
\end{equation}
which sets the relation between the measured (real) quantity and the abstract definition of (complex) weak values.
Consulting Eq.~\eqref{eq:WVdefinitionMZI} we conclude that we can conveniently get rid of the factor $1/\cS$ by defining the normalized WV,
\begin{equation}
\overline{\av{Q_2}}^M_{WV}\eqd \frac{\av{Q_2}^M_{WV}}{\av{Q_2}^M_{WV}\at{11}}.
\label{eq:NormalizedWV}
\end{equation}
Here $\av{Q_2}^M_{WV}\at{11}$ is the measured $\av{Q_2}_{WV}$, when both QPCs A and B are set to full transmission. For this tuning, it follows that $\av{Q_2}^M_{WV}\at{11}=\av{Q_2}_{WV}=\av{Q_2}_{max}$.
Eq. (\ref{eq:NormalizedWV}) involves only measurable quantities and serves to operatively identify weak values beyond the range allowed by unconditioned measurements.

\begin{figure}
	\includegraphics[width=8.6cm]{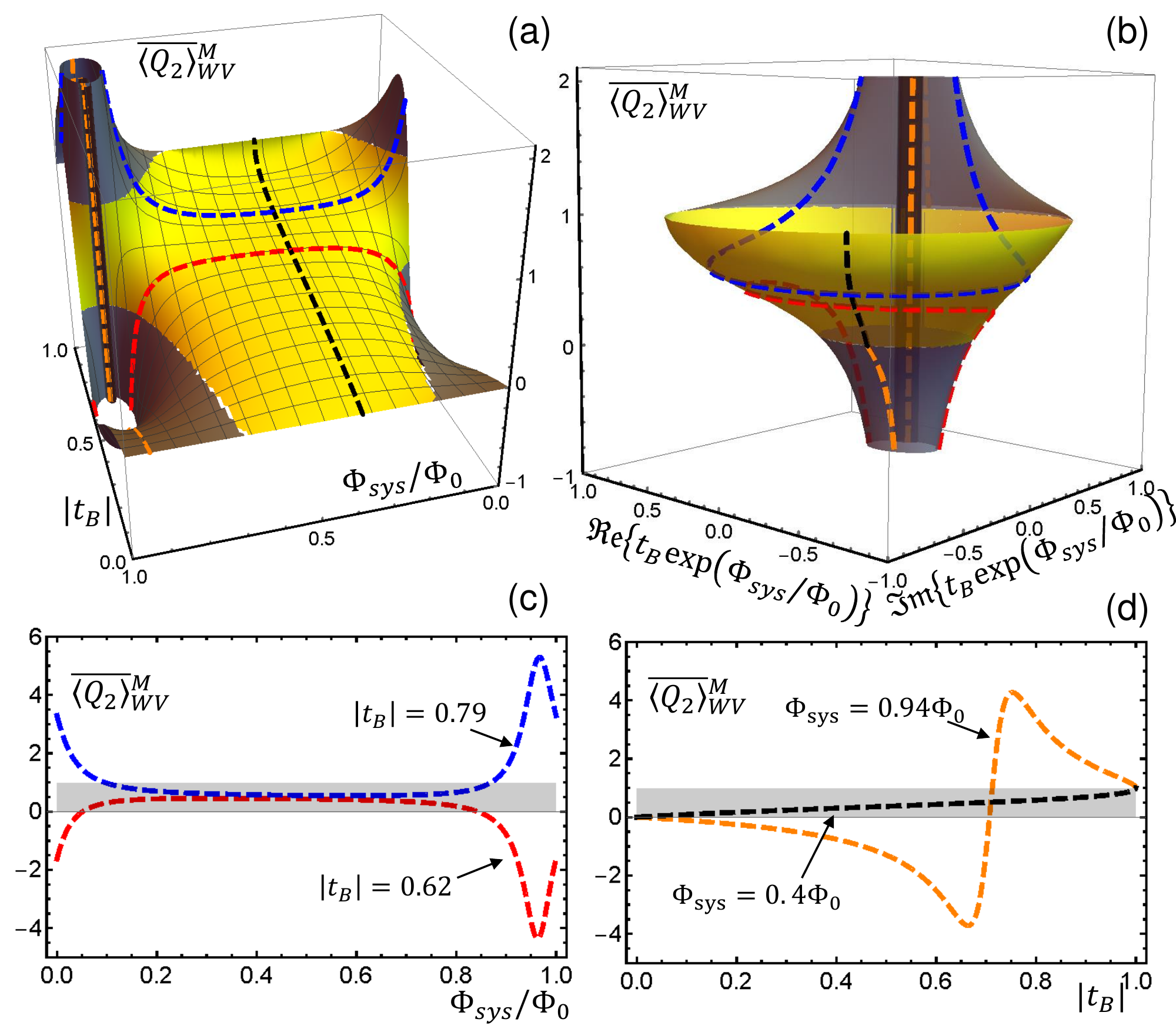}\\
	\caption{(Color online) The normalized WV, $\overline{\av{Q_2}}^M_{WV}$, as a function of transmission amplitude of QPC B, $\abs{t_B}$, and the flux, $\F_{sys}/\F_0$. The transmission of QPC A is set to $\abs{t_A}=\frac{1}{\sqrt{2}}$, and $\tilde\lm=0.05\p$. An overall phase $\dl=0.1\p$, representing the asymmetry of the interferometer's arms is included, implying that the interference pattern is not symmetric around $\F_0/2$.
		(a). A 3D plot of $\overline{\av{Q_2}}^M_{WV}$ as a function of $t_B$ and $\F_{sys}/\F_0$. (b). $\overline{\av{Q_2}}^M_{WV}$ as a function of the real and the imaginary parts of $t_B \exp{(\F_{sys}/\F_0)}$. The yellow colored region corresponds to values that fall within the conventional range, $[0,1]$, while the gray regions underline the observability of ``exceptional values'' that lie beyond this range. The dashed lines correspond to specific cuts along $\abs{t_B}=0.62$ (red), $\abs{t_B}=0.79$ (blue), $\F_{sys}=0.94\F_0$ (orange), and $\F_{sys}=0.4\F_0$ (black). Those are depicted in (c) vs. $\F_{sys}/\F_0$ and in (d) vs. $\abs{t_B}$, with the same respective colors.\label{fig:Results}}
\end{figure}

\section{Results and discussion}

The present analysis is aimed at implementing the general framework of WV protocol \cite{Aharonov1988} to a representative electronic system. The latter is experimentally accessible \cite{Weisz2014}, rendering the present
protocol amenable to experimental verification. We have focused on one central aspect of WV, namely  the possibility to obtain expectation values (a.k.a. \ti{weak values}) that lie beyond the range of possible outcomes of strong measurement \cite{Neumann1955} (the latter implies the collapse of the system's wave function). Specifically, our setup consists of a ``system'' and a ``detector'' (\MZIsys and \MZIdet, cf. Fig.~\ref{fig:Setup}) which are electrostatically weakly coupled. The detector is tuned to measure the charge transmitted through one of the system interferometer's arms (a weak ``which path'' measurement \cite{Aleiner1997,Levinson1997,Gurvitz1997,Buks1998}).

Intuition based on strong measurement procedure would suggest that when one electron is injected into the system's MZI,  the normalized charge that can be measured on one of the interferometer arms is anything between 0 and 1. By contrast, WV protocol allows us to obtain values which are above this value (``charge larger than 1'' or even \ti{negative}). The results shown in Fig.~\ref{fig:Results} make it clear that as far as weak values are concerned, both conventional values (that conform to ``allowed values'' of strong measurement) \ti{and} exceptional values which lie beyond the interval $[0,1]$ are possible.

Specifically consider Figures~\ref{fig:Results}a and \ref{fig:Results}b. We show that $\overline{\av{Q_2}}^M_{WV}$ may grow beyond the range of values allowed by a strong measurement of $Q_2$, which involves the collapse of the system's wavefunction. This is highlighted by a different (gray) color of the 3D plot. Figures~\ref{fig:Results}c and \ref{fig:Results}d show $\overline{\av{Q_2}}^M_{WV}$ along specific cuts of the 3D plot, for $\abs{t_B}=0.62$ and $\abs{t_B}=0.79$ as a function of $\F_{sys}/\F_0$, and $\F_{sys}=0.94\cdot\F_0$ and $\F_{sys}=0.4\cdot\F_0$ as a function of $\abs{t_B}$. We note that the WV may become exceedingly large (positive or negative) for a proper choice of the parameters.

In the procedure outlined above, we have put special emphasis on how the measured values of current and current correlations should be calibrated to fit with the weak value formalism. It would be interesting to repeat this analysis having in mind variations of our setup (e.g., replacing the MZI detector by a QPC or a current carrying quantum dot).

\begin{acknowledgements}
	We acknowledge discussions with H. Choi, M. Heiblum, I. Sivan, and E. Weisz. This work has been supported by Deutsche Forschungsgemeinschaft (DFG) grant RO 2247/8-1, SH 81/3-1, and RO 4710/1-1, the Israel Science Foundation (ISF), the Minerva foundation, and the Russia-Israel IMOS project.
\end{acknowledgements}

\appendix

\section{Appendix 1: Explicit expressions for electronic trajectories}\label{sec:ExplicitTrajectories}
\subsection{Solution to the single particle Hamiltonian}
\begin{figure}
	\includegraphics[width=8.6cm]{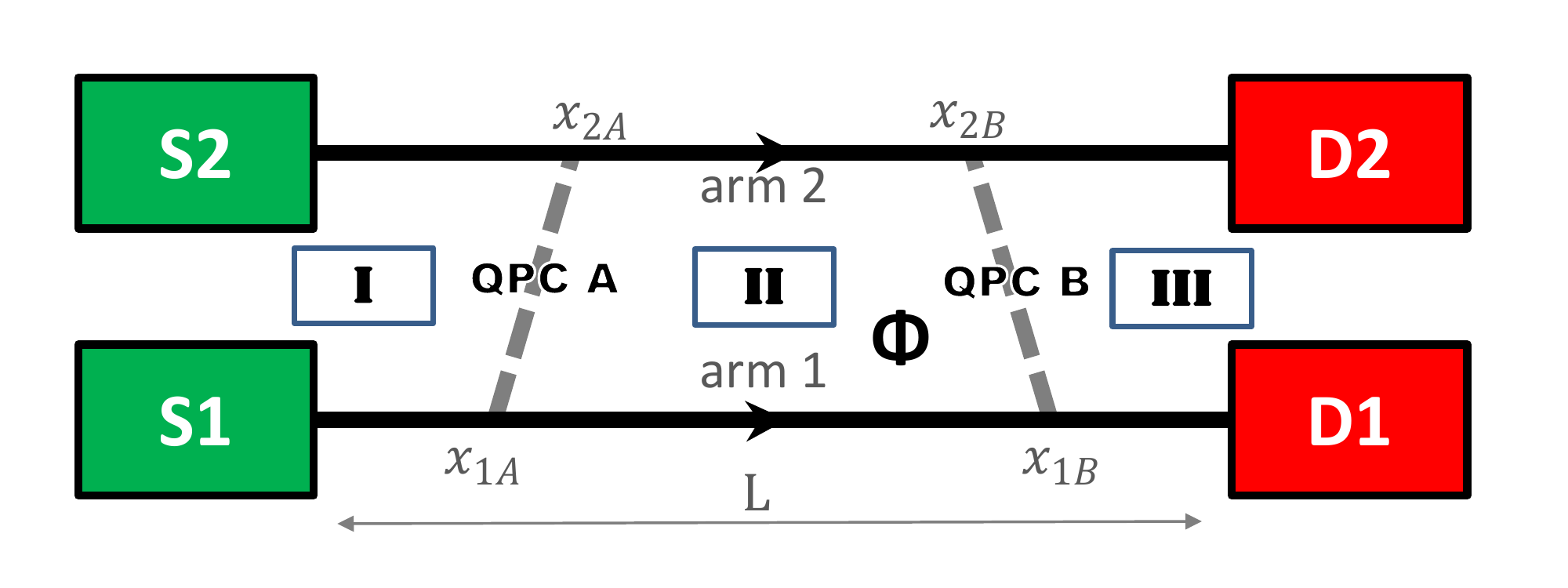}\\
	\caption{(Color online) A schematics of a MZI. The points $x_{1\a}$ and $x_{2\a}$, $\a=A,B$ at QPCs A and B are connected by the tunneling terms $\G_\a$. The Latin numerals denote the various sectors of the MZI: I - left to QPC A, II - between QPC A and B, III- right of QPC B. \label{fig:SetupAppendix}}
\end{figure}

Here we solve the single-particle problem for a single MZI (suppose \MZIsys from figure \ref{fig:Setup}) and expand the solution in partial electronic trajectories (cf. Fig.~\ref{fig:Paths}). We begin with the single particle Schr\"odinger equation
\begin{equation}
\bR{i\hb\frac{\dpa}{\dpa t}-\hat H}\Y=0
\end{equation}
where $\Y=\mat{\y_1(x,t)\\ \y_2(x,t)}$, and
	\begin{equation}
	\hat H\Y =\mat{iv\dx\y_1(x,t)+\sum_{\a=A,B}\bS{\frac{\G_\a\as}{2}\dl(x-x_{1\a}^+)\y_2(x_{1\a}^-,t)+\frac{\G_\a\as}{2}\dl(x-x_{1\a}^-)\y_2(x_{1\a}^+,t)}\\
		iv\dx\y_2(x,t)+\sum_{\a=A,B}\bS{\frac{\G_\a}{2}\dl(x-x_{2\a}^+)\y_1(x_{2\a}^-,t)+\frac{\G_\a}{2}\dl(x-x_{2\a}^-)\y_1(x_{2\a}^+,t)}}.
	\label{eq:ScatteringHamiltonian}
	\end{equation}
Here $\y_1(x,t)$ and $\y_2(x,t)$ denote the wavefunctions in the corresponding arms 1 and 2, $\G_\a$ represents the tunneling term associated with the $\a$-th QPC, connecting points $x_{1\a}$ and $x_{2\a}$, $\a=A,B$ (cf. Fig.~\ref{fig:SetupAppendix})); $x^{\pm}= \lim_{\ve\to0}x\pm \ve$. $\G_\a$ may be related to the scattering amplitudes through Eqs.~\eqref{eq:TunnelingAmplitudesAppendix}.

This problem is diagonal in the scattering basis
\begin{equation}
\Y_{k,l}(x,t)=\frac{1}{\sqrt L}e^{ik(x-vt)}\condf{
	\vec \n_l&,x\in \text{I}\\
	\breve S_A\vec \n_l&,x\in \text{II}\\
	\breve S_B\vec \n_l&,x\in \text{III}
}.
\label{eq:ScatteringStates}
\end{equation}
Here the Latin numerals denote the various sectors of the MZI: I - left to QPC A, II - between QPC A and B, III- right of QPC B, $\breve S_\a=\mat{r_\a&-t_\a\as\\ t_\a&r_\a}$ is the scattering matrix at QPC $\a$ and the scattering amplitudes are
\begin{subequations}
	\begin{eqnarray}
	r_\a=\frac{(2v^2)-\abs{\G_\a}^2}{(2v^2)+\abs{\G_\a}^2}\\
	t_\a=\frac{4iv\G_\a}{(2v^2)+\abs{\G_\a}^2},
	\end{eqnarray}\label{eq:TunnelingAmplitudesAppendix}
\end{subequations}
for the symmetric MZI case ($x_{1B}-x_{1A}=x_{2B}-x_{2A}$). The index $l=1,2$ denotes two orthogonal solutions $\vec \n_1=\mat{1\\0}$ and $\vec \n_2=\mat{0\\1}$ that correspond to the scattering state incident from $S1$ or $S2$.

\subsection{Explicit expressions for the coefficients $c_i$}
The probability of the particle incident from $S1$ to be detected in $D1$ can be presented using the path integral formalism as
\begin{equation}
P_{S1\to D1}=\abs{\int_\cC\cD \Y e^{iS\bC{\Y}/\hb}}^2,
\end{equation}
where $\cC$ represents all the trajectories from $S1$ to $D1$, and $c_i\eqd e^{iS\bC{\Y_i}/\hb}$ is the weight of the corresponding trajectory. The same argument may be repeated for probabilities $P_{S1\to D2}$, $P_{S2\to D1}$, and $P_{S2\to D2}$ to include all the trajectories depicted in Fig.~\ref{fig:Paths}.
The explicit expressions for the trajectory weights, $c_i$,  may be found from the exact solution, Eq.~\eqref{eq:ScatteringStates}. Here we summarize the results. Up to unimportant orbital phases
\begin{subequations}
	\begin{eqnarray}
	c_1^{sys}&=&-t_A\as r_B\\
	c_2^{sys}&=&r_A r_B\\
	c_3^{sys}&=&-r_A t_B\as\\
	c_4^{sys}&=&-t_A\as t_B\\
	c_1^{det}&=&-t_C t_D\as\\
	c_2^{det}&=&r_C t_D\\
	c_3^{det}&=&r_C r_D\\
	c_4^{det}&=&t_C r_D.
	\end{eqnarray}
\end{subequations}

\section{Appendix 2: Derivation of expectation values and correlators \label{sec:DerivationCorrelators}}
Here we derive explicit expressions for expectation values of operators that appear throughout the manuscript.
For simplicity we assume a symmetric MZI ($L_1=L_2$, $L_3=L_4$) operating in the low frequency, zero temperature regime, where the quantum state does not vary much over the time of the experiment, hence is almost steady. Thus, in this regime, all quantities are essentially time independent. We evaluate the expectation values by computing a trace of the operator with respect to the initial density matrix, $\ro_i=\ketbra{S1,S4}{S1,S4}$ describing two particles, which are taken fom out-of equilibrium distribution,and which are incident from the biased sources $S1$ and $S4$ (cf. Fig.~\ref{fig:Setup}). At the end of this appendix (cf. Section~\ref{subsec:BackgroundChargeAppendix}) we add a contribution from the equilibrium, background sea of electrons below the Fermi level. Throughout this section, the charge operators associated with each MZI are normalized to have no physical units. The latter may be recovered at the end, multiplying the normalized charge by $\av{Q_i}_0$ ($i=2,3$) (those are defined following Eqs.~\eqref{eq:Q2WVDefinition} and \eqref{eq:Q3WVDefinition}).

\subsection{Derivation of equations~\eqref{eq:Q2WVDefinition} and \eqref{eq:Q3WVDefinition}}
Writing explicitly the expectation values appearing in eq.~\eqref{eq:WVdefinition}, we come up with the expression
\begin{equation}
\av{Q_2}_{WV}=\frac{\braoket{S1,S4}{Q_2 I_{D1}}{S1,S4}}{\braoket{S1,S4}{I_{D1}}{S1,S4}},
\end{equation}
where $I_{D1}=\ketbra{D1}{D1}$. To leading order in the coupling, $\lm$ (cf. eq.~\eqref{eq:PartialPaths}), the two MZIs may be considered as decoupled, and the trace over \MZIdet states ($\ketbra{S4}{S4}$) is trivial. Hence the expression reads
\begin{equation}
\av{Q_2}_{WV}=\frac{\braoket{S1}{Q_2}{D1}}{\braket{S1}{D1}}.
\label{eq:Q2WVAppendix}
\end{equation}
The operator $Q_2$ is proportional to a projection operator that selects only partial wavepackets $\ket{\y_{sys}^1}$ and $\ket{\y_{sys}^4}$ (cf. Fig.~\ref{fig:Paths}). Only $\ket{\y_{sys}^1}$ has support at $D1$ and contributes to the numerator. The denominator includes two contributions, $\ket{\y_{sys}^2}$ and $\ket{\y_{sys}^4}$. It follows that
\begin{equation}
\av{Q_2}_{WV}=\frac{c^{sys}_4}{c^{sys}_4+c^{sys}_2}.
\end{equation}
The derivation of an expression for $\av{Q_3}_{WV}\eqd\frac{\av{Q_3\cdot I_{D4}}}{\av{I_{D4}}}$ (eq.~\eqref{eq:Q3WVDefinition}) is similar.

\subsection{Expression for the system--detector current correlator\label{subsec:CurCurCorrelator}}
Here we derive the expression for $\av{I_{D1}I_{D4}}$ to leading order in the coupling $\lm$. The expression for the non-equilibrium currents correlator reads
\begin{equation}
\av{I_{D1}I_{D4}}=\braoket{S1,S4}{I_{D1}I_{D4}}{S1,S4}.
\end{equation}
We plug in the explicit expressions for the currents, $I_{D1}=\ketbra{D1}{D1}$ and $I_{D4}=\ketbra{D4}{D4}$ to obtain,
\begin{equation}
\av{I_{D1}I_{D4}}=\abs{\braket{S1,S4}{D1,D4}}^2.
\label{eq:CurCurCorrAppendix}
\end{equation}
We note that due to charge conservation and the assumption of a steady state, the sum of the currents on the two arms (between the QPCs) of any MZI, is equal to the total current that flows into the MZI. For example, for \MZIsys,
\begin{equation}
I_{S1}=I_1+I_2,
\label{eq:CurrentConservation1Appendix}
\end{equation}
where $I_{S1}$ denotes an operator that measures current flow through $S1$, and $I_{1(2)}$ measure currents at arbitrary points on arm 1 (2) (cf. Fig~\ref{fig:Setup}). Next, we integrate Eq.~\eqref{eq:CurrentConservation1Appendix} over the time-of flight of electron in the MZI. Integration of $I_{S1}$ over this time yields the total charge that flows into the MZI during the time-of-flight. The latter is noiseless according to our assumptions and therefore is proportional to the identity (exactly one electron has been injected from $S1$). The integration of $I_{1(2)}$ yields the fraction of charge that when to arm 1(2), $Q_{1(2)}$. It follows from the above that
\begin{equation}
Q_1+Q_2=\I,
\label{eq:ChargeConservation1Appendix}
\end{equation}
where we have used $\ta_{fl}I_{S1}=\I$ in dimensionless units. A similar identity may be obtained for \MZIdet,
\begin{equation}
Q_3+Q_4=\I.
\label{eq:ChargeConservation2Appendix}
\end{equation}
Next, we insert those unit operators into Eq.~\eqref{eq:CurCurCorrAppendix}, which yields the expression
\begin{equation}
\begin{split}
\av{I_{D1}I_{D4}}=\Big|\bra{S1,S4}&Q_1 Q_3+ Q_1 Q_4+\\
+ &Q_2 Q_3+Q_2 Q_4\ket{D1,D4}\Big|^2.
\end{split}
\end{equation}
Each element in the sum is diagonal in the basis of trajectories, and can be evaluated employing the wavefunction \eqref{eq:PartialPaths}. It follows that
\begin{equation}
\begin{split}
\av{I_{D1}I_{D4}}=\Big|&\av{Q_1}_{S1;D1} \av{Q_3}_{S4;D4}+ \av{Q_1}_{S1;D1}\av{Q_4}_{S4;D4}+\\
+ &e^{i\lm}\av{Q_2}_{S1;D1}\av{Q_3}_{S4;D4}+\av{Q_2}_{S1;D1}\av{Q_4}_{S4;D4}\Big|^2,
\end{split}
\end{equation}
where we have introduced the notation $\av{Q}_{S;D}=\braoket{S}{Q}{D}$. Using the identities \eqref{eq:ChargeConservation1Appendix} and \eqref{eq:ChargeConservation2Appendix}, and expanding to leading order in $\lm$, we arrive at the expression
\begin{equation}
\begin{split}
\av{I_{D1}I_{D4}}=\Big|&\av{\I}_{S1;D1} \av{\I}_{S4;D4}+i\lm\av{Q_2}_{S1;D1}\av{Q_3}_{S4;D4}\Big|^2.
\end{split}
\end{equation}
Simplifying it, employing the relations $\av{I_{D1}}_0\eqv\abs{\av{\I}_{S1;D1}}^2$, $\av{I_{D4}}_0\eqv\abs{\av{\I}_{S4;D4}}^2$ (see for example Eq.~\eqref{eq:CurCurCorrAppendix} and eq.~\eqref{eq:Q2WVAppendix}), we obtain
\begin{equation}
\begin{split}
\av{I_{D1}I_{D4}}=\av{I_{D1}}_0\av{I_{D4}}_0\bR{1+2\lm\im{\av{Q_2}_{WV}\av{Q_3}_{WV}}}.
\label{eq:CurCurCorrelatorAppendix}
\end{split}
\end{equation}

\subsection{Expressions for the current expectation values}
Here we compute the expressions for the current expectation values. To do so, we employ the charge conservation in a steady state regime. We repeat the discussion following Eq.~\eqref{eq:ChargeConservation1Appendix} that led to the identities
\begin{subequations}
	\begin{eqnarray}
	I_{D1}+I_{D2}&=&\I\\
	I_{D3}+I_{D4}&=&\I.
	\end{eqnarray}\label{eq:ChargeConservation3Appendix}
\end{subequations}
The latter may be employed to write
\begin{subequations}
	\begin{eqnarray}
	\av{I_{D1}}&=&\av{I_{D1}I_{D3}}+\av{I_{D1}I_{D4}}\\
	\av{I_{D4}}&=&\av{I_{D2}I_{D4}}+\av{I_{D1}I_{D4}}.
	\end{eqnarray}\label{eq:CurrentAverageIdentityAppendix}
\end{subequations}
Since $\av{I_{D1}I_{D4}}$ was found in eq.~\eqref{eq:CurCurCorrelatorAppendix}, one may follow the derivation in appendix~\ref{subsec:CurCurCorrelator} to obtain an expression for $\av{I_{D1}I_{D3}}$ and $\av{I_{D2}I_{D4}}$. Plugging those results into Eq.~\eqref{eq:CurrentAverageIdentityAppendix} one ends up with the equalities
\begin{subequations}
	\begin{eqnarray}
	&&\label{eq:CurrentNECorrelatorID1Appendix}\av{I_{D1}}=\av{I_{D1}}_0\bR{1+2\lm\im{\av{Q_2}_{WV}\av{Q_3}}}\\
	&&\av{I_{D4}}=\av{I_{D4}}_0\bR{1+2\lm\im{\av{Q_3}_{WV}\av{Q_2}}}.
	\end{eqnarray}\label{eq:CurrentNECorrelatorAppendix}
\end{subequations}

\subsection{The contribution of the background charge\label{subsec:BackgroundChargeAppendix}}
Above we have derived expressions for the average currents and current-current correlator in the presence of single particle taken from out-of-equilibrium distribution. In real life, there is background charge. The latter may interact with the incoming electrons, producing a shift in the measured signal. We consider the background charge as a noiseless constant that shifts the charge operator on each arm ($Q_2\to Q_2+\av{Q^{bg}_2}$, $Q_3\to Q_3+\av{Q^{bg}_3}$). We may now rewrite now eqs.~\eqref{eq:CurCurCorrelatorAppendix} and \eqref{eq:CurrentNECorrelatorAppendix} with the contribution of the background charge,

\begin{subequations}
	\begin{eqnarray}
	&&\av{I_{D1}}=\av{I_{D1}}_0\bR{1+2\lm\im{\av{Q_2}_{WV}\bS{\av{Q_3}+\av{Q_3^{bg}}}}}\\
	&&\av{I_{D4}}=\av{I_{D4}}_0\bR{1+2\lm\im{\av{Q_3}_{WV}\bS{\av{Q_2}+\av{Q_2^{bg}}}}}
	\end{eqnarray}
\end{subequations}

\begin{equation}
\av{I_{D4}I_{D1}}=\av{I_{D4}}_0\av{I_{D1}}_0\bS{1+2\lm \im{\bS{\av{Q_2}_{WV}+\av{Q_2^{bg}}}\bS{\av{Q_3}_{WV}+\av{Q_3^{bg}}}}}
\end{equation}
(cf. Eq.~\eqref{eq:DetectorCurrent}).

\bibliographystyle{aps-nameyear}

\end{document}